# A Composite Fault Diagnosis Model for NPPs Based on Bayesian-EfficientNet Module

Siwei Li , Jiangwen Chen, Hua Lin, Wei Wang

***Abstract*—This article focuses on the faults of important mechanical components such as pumps, valves, and pipelines in the reactor coolant system, main steam system, condensate system, and main feedwater system of nuclear power plants (NPPs). It proposes a composite multi-fault diagnosis model based on Bayesian algorithm and EfficientNet large model using data-driven deep learning fault diagnosis technology. The aim is to evaluate the effectiveness of automatic deep learning-based large model technology through transfer learning in nuclear power plant scenarios.**

***Keywords—Nuclear Power Plant, Fault Diagnosis, Deep learning, Large Model***

## I. Introduction

The primary and secondary circuits are the core components of nuclear power units, with high frequency of faults, wide impact range, and high diagnostic difficulty. The traditional nuclear power fault diagnosis method analyzes the collected data and identifies potential problems or abnormal patterns through statistical, trend analysis, spectrum analysis, and other methods. This method relies on expert experience and lacks adaptability and generalization ability. In recent years, machine learning has developed rapidly, and Artificial Neural Network (ANN) networks have been applied to NPPs fault diagnosis[1]. However, the training process requires manual adjustment of network weights and bias parameters, and its performance depends on multiple factors such as network structure, training data, and optimization algorithms, resulting in limited generalization ability[2]. With the breakthrough progress of large models in deep learning in complex system modeling and pattern recognition, large models combined with parameter optimization algorithms will be applied to the diagnosis of faults in the primary and secondary circuits of NPPs. By constructing a highly robust and strong generalization diagnostic model, the speed, accuracy, and interpretability of fault diagnosis in NPPs can be improved, providing guarantees for safe and efficient operation. Therefore, conducting research on fault diagnosis methods based on deep learning for the primary and secondary circuits of NPPs is of great significance.

Currently, most of the proposed large-scale models are developed around areas such as image classification and segmentation, object monitoring, text classification, video classification, audio classification, and machine translation[3]. Experience has shown that large models can perform reasonable calculations on classification and regression problems, but they have not yet delved into the field of industrial control, so they still have great prospects in the field of NPP fault diagnosis.

## II. Methods

### A. *Bayesian Optimization Algorithm*

For neural networks, setting hyperparameters is crucial for learning outcomes. The widely used search hyperparameter methods include manual parameter tuning, Grid Search, Randomized Search, Bayesian Optimization, and so on. The parameters of large models are exceedingly vast, and the network structure is highly intricate, rendering it arduous to acquire effective hyperparameters through manual parameter tuning. The grid search method is an exhaustive method in the search space and is not suitable for large hyperparameter searches. Random search is faster than network search, but the combination of hyperparameters may result in worse results. Therefore, we choose Bayesian optimization method.

Bayesian optimization was developed by J Snoek , and its main idea is to update the posterior distribution of the optimized objective function (a generalized function that only requires specifying inputs and outputs, without knowing the internal structure and mathematical properties) by continuously adding sample points until the posterior distribution basically fits the true distribution[4].

In the mathematical process of Bayesian optimization algorithm, we mainly perform the following steps: (1) Define the domain of $f(x)$ and x that need to be estimated; (2) Take out a finite number of n values on $x$, and solve for the corresponding $f(x)$ of these $x$ values (by solving the observed values); (3) Based on limited observations, estimate the function (this assumption is referred to as prior knowledge in Bayesian optimization) and obtain the target value (maximum or minimum) of the estimate on $f^*(x)$. (4) Define a certain rule to determine the next observation point to be calculated.

### *B. EfficientNet Large Model*

Large models usually refer to a neural network model with vast of parameters, high computational complexity, and complex model structure. The purpose of its design is to improve the expressive and predictive performance of the model, enabling it to handle more complex tasks and data[5].

Large models can be mainly divided into language models, visual models, and multimodal models. As is well known, GPT (Generative Pre-trained Transformer) is a pre-trained language model based on the Transformer architecture. Due to the characteristics of fault diagnosis tasks, we choose to conduct research based on visual large models. ViT (Vision Transformer), SE ResNeXt, Inception, EfficientNet, etc. are widely used models in CV large models. EfficientNet applies an efficient neural network architecture, which achieves efficient model design by uniformly scaling the depth, width, and resolution of the network, thereby reducing computation and parameter complexity while maintaining accuracy[6]. It is a high-performance solution designed specifically for image classification and recognition when computing resources are limited. Therefore, we decide to compare the EfficientNet model with some commonly used large models to explore the application prospects of EfficientNet in the field of nuclear power plant fault diagnosis.

However, large models require a large amount of computing resources and large-scale data for training, and for small and medium-sized research institutes or institutions, the cost of developing or training their own large models from scratch is difficult to accept. This poses difficulties for the deployment of large models in specific industrial fields. For the above reasons, we chooses to deploy the pre-trained large model EfficientNet and other models through transfer learning, which can also provide a feasible reference for other researchers to use large models[7].

The overall algorithm flow is shown in **Fig.1.** The normalized data is processed into images using the grayscale algorithm, and then 70% of the training set image data is used to search for model hyperparameters. The resulting hyperparameters are fed into EfficientNet along with the training data.

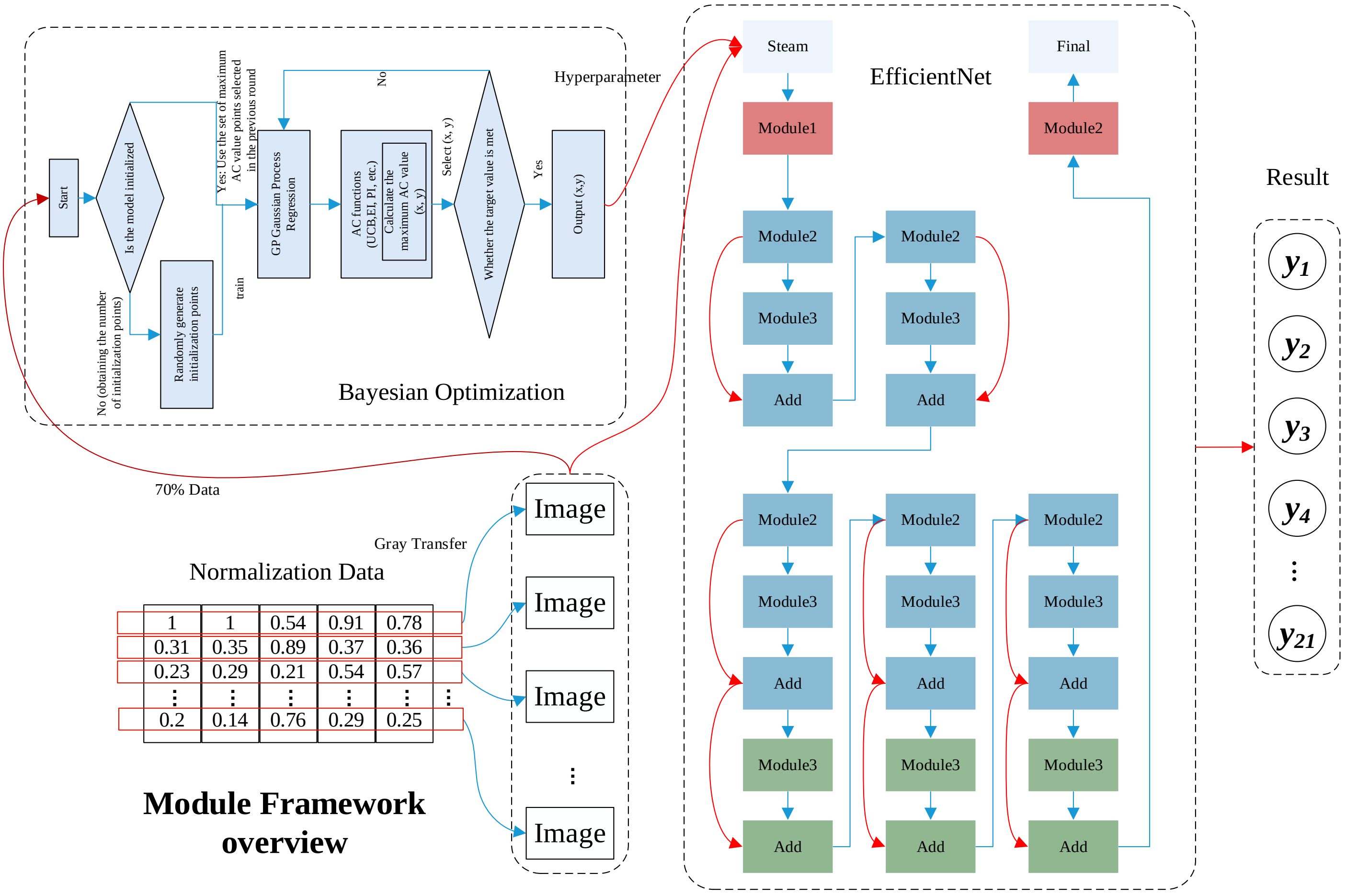

Fig 1. Module framework overview

## III. EXPERIMENT AND RESULTS

### A. Dataset description

We conducted simulation modeling on AP1000 units and ran the simulation models in three power scenarios: 50% FP (Full Power), 75% FP, and 100% FP. Inject 6 typical faults into the model uniformly in each power scenario and collect data from 10725 real-time parameters at 1200 time points under the faults. In addition, for the 100% FP power scenario, three additional composite operating conditions are injected. (as shown in **Table I** ).

TABLE I. FAULT CONDITIONS AND CORRESPONDING POWER SCENARIOS

| ***Number*** | ***Fault name*** | ***Power scenario*** |
|---|---|---|
| 0 | Reduced coolant inventory in the primary circuit | 50%FP、75%FP、100%FP |
| 1 | Increased coolant inventory in the primary circuit | 50%FP、75%FP、100%FP |
| 2 | Increased heat dissipation in the secondary circuit | 50%FP、75%FP、100%FP |
| 3 | Reduced heat dissipation in the secondary circuit | 50%FP、75%FP、100%FP |
| 4 | Reactive accidents | 50%FP、75%FP、100%FP |
| 5 | Coolant loss of flow | 50%FP、75%FP、100%FP |
| 7 | Cold pipe break superimposed with steam pipe break | 100%FP |
| 8 | Cold pipe break superimposed with main water supply pipe break | 100%FP |
| 9 | Breaks in the main water supply pipeline and stacking of control rods | 100%FP |

*B.     Data Pre-processing*

The raw data collected are dimensional with significant differences in magnitude. Consequently, it is necessary to normalize the original data by removing dimensions and converting them to the same level. In addition, if we can ensure a stable distribution of nonlinear inputs during training, we can reduce problems such as vanishing gradients due to saturation and thus speed up network training[8]. So we use Min-max normalization to normalize the data, and the normalized data range is$[0,1]$.

$$x' = \frac{x - \min(x)}{\max(x) - \min(x)} \quad (1)$$

Among them, $\min(x)$ 、 $\max(x)$ is the minimum and maximum values of the sample data, respectively.

The collected data consists of over 200 million data points. Directly inputting them into the model for training not only yields poor performance but also poses challenges in the training process.  The fault data from the simulation model is typical of multiple sets of one-dimensional time series data, and the correlation and characteristics between the data are not obvious. Transforming one-dimensional data into two-dimensional data can make it easier to identify repetitive patterns and abnormal features in images. The commonly used methods for two-dimensional data include wavelet transform (WT), Grami angle field method (GAFS), Markov transition field method (MTF) and so on[9].

The methods analyze the temporal evolution of a single parameter in the time dimension, providing insights into its periodicity and autocorrelation characteristics. However, they fail to capture the comprehensive representation and overall characteristics of nuclear power plant parameters.

Based on the above analysis, we decide to use image grayscale algorithm to process all the raw data (i.e.10725 parameter data) collected at each time step of the model into a $104 * 104$ two-dimensional grayscale image to obtain a two-dimensional representation of the overall parameters of the nuclear power plant. After processing by the image grayscale algorithm, the data mapping results are shown in **Fig 2**.

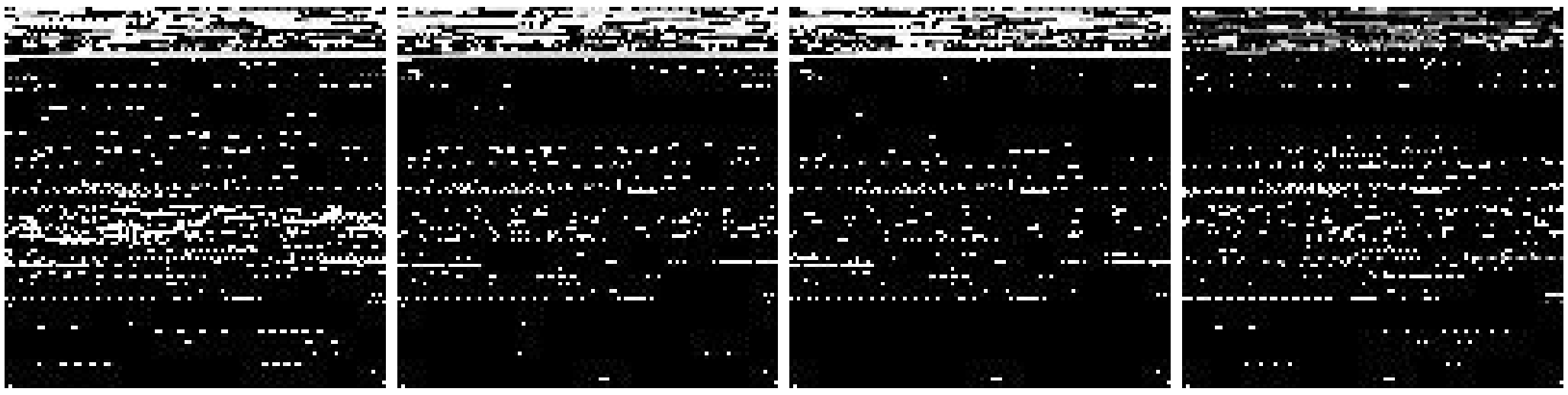

Fig 2. Grayscale visualization of overall data of nuclear power plants

TABLE II.    HYPERPARAMETER OPTIMIZATION RESULTS

| ***Hyperparameter*** | Value | ***Hyperparameter*** | Value |
|---|---|---|---|
| ***Batch Size*** | 32 | ***Use Warm Up*** | true |
| ***Epochs*** | 100 | ***Warm Up Steps*** | 500 |
| ***Weight Decay*** | L2 | ***Learning Rate*** | 1.98E-3 |
| ***Optimizer*** | Momentum | ***Anchor_Ratio*** | 3 |
| ***Momentum*** | 0.9 | ***Anchor Scale*** | 1 |

### C. Bayesian Optimization

We set the initial number of Bayesian search points to 10, and simultaneously run 5 Bayesian search algorithms. 70% of the training set data is randomly sampled for Bayesian search. When the accuracy of the search results reaches 90%, the search is stopped in advance and the hyperparameter combination is returned. The search results of hyperparameter combination are shown in **Table II**.

### D. Results and comparison

We will input the hyperparameters obtained from the Bayesian Optimization algorithm into the EfficientNet model for training. And by comparing the diagnostic results of the proposed method with other widely used large models on the same dataset for multiple fault scenarios of multiple power single faults and full power composite faults in nuclear power plant AP1000 units (as shown in **Table III**), the large model can correctly identify most of the confusing nuclear power plant fault images for professionals. The accuracy represents the proportion of the number of correctly diagnosed faults to the total number of sample faults. As we can see from the table, the accuracy of the model in diagnosing up to 21 types of faults in the test set was the lowest at 91.5% and the highest at 96.4%. Generally, a larger F1-Score value indicates a higher quality model. It is the harmonic mean of precision and precision and interacts with recall. The EfficientNet model achieves 95.3% F1-Score. Precision is the fraction of all examples that the model predicts to be positive that are actually positive, and it is concerned with how well the model predicts a positive example.

Recall is the fraction of all positive examples that are predicted to be true. As with precision, higher is better. Overall, the predicted results of the model are consistent with the actual fault types. The large-scale model method based on automatic deep learning can effectively improve the efficiency of fault diagnosis in nuclear power plants.

EfficientNet and other large models encounter difficulties in identifying highly similar faults, possibly due to the presence of numerous redundant parameters within the overall 10725 parameters collected from the nuclear power plant at each moment. These redundancies occupy a significant portion of the grayscale feature image and are mistakenly recognized as important features by the model, leading to reduced prediction accuracy and potential deviations. Humans can effectively identify power plant faults based on composite information such as displayed images and sounds. However, large models represented by EfficientNet only rely on images, and their recognition accuracy is close to or even higher than that of professionals. This once again highlights the broad application prospects of large model technology based on automatic deep learning in the field of nuclear power plant fault diagnosis.

TABLE III. COMPARISON OF DIAGNOSTIC RESULTS AMONG MAJOR MODELS

| ***Model*** | ***Accuracy*** | ***F1-Score*** | ***Precision*** | ***Recall*** |
|---|---|---|---|---|
| InceptionV4 | 0.915 | 0.914 | 0.948 | 0.915 |
| SE ResNeXt50 | 0.919 | 0.921 | 0.940 | 0.919 |
| Xception71 | 0.926 | 0.941 | 0.950 | 0.926 |
| EfficientNetB0 | 0.954 | 0.953 | 0.964 | 0.954 |

## IV. SUMMARY

In response to the shortcomings of traditional nuclear power fault diagnosis methods, this paper proposes a large model combining parameter optimization algorithms, aiming to improve and optimize the efficiency of nuclear power fault diagnosis. The experimental results show that the fault diagnosis model based on Bayesian algorithm and EfficientNet large model exhibits high accuracy in diagnosing single or composite faults of main components in the primary and secondary circuits of nuclear power systems at different power levels. This discovery fully demonstrates the enormous potential and feasibility of large-scale deep learning models in fault diagnosis in the field of nuclear power, providing strong technical support for safe operation of nuclear power.

## ACKNOWLEDGMENT

This research did not receive any specific grant from funding agencies in the public, commercial, or not-for-profit sectors.

## REFERENCES

[1] W. Rawat and Z. Wang, "Deep Convolutional Neural Networks for Image Classification: A Comprehensive Review," Neural Computation, vol. 29, no. 9, pp. 2352-2449, 2017, doi: 10.1162/neco_a_00990.

[2] X. Li, K. Cheng, T. Huang, Z. Qiu, and S. Tan, "Research on short term prediction method of thermal hydraulic transient operation parameters based on automated deep learning," Annals of Nuclear Energy, vol. 165, p. 108777, 2022/01/01/ 2022, doi: https://doi.org/10.1016/j.anucene.2021.108777.

[3] Y. Quéau, F. Leporcq, A. Lechervy, and A. Alfalou, Learning to classify materials using Mueller imaging polarimetry. 2019, p. 1.

[4] J. Snoek, H. Larochelle, and R. P. Adams, "Practical Bayesian optimization of machine learning algorithms," presented at the Proceedings of the 25th International Conference on Neural Information Processing Systems - Volume 2, Lake Tahoe, Nevada, 2012.

[5] J. Yang et al., Harnessing the Power of LLMs in Practice: A Survey on ChatGPT and Beyond. 2023.

[6] M. Tan and Q. V. Le, "EfficientNet: Rethinking Model Scaling for Convolutional Neural Networks," ArXiv, vol. abs/1905.11946, 2019.

[7] J. Kaddour, J. Harris, M. Mozes, H. Bradley, R. Raileanu, and R. McHardy, "Challenges and Applications of Large Language Models," ArXiv, vol. abs/2307.10169, 2023.

[8] S. Ioffe and C. Szegedy, "Batch Normalization: Accelerating Deep Network Training by Reducing Internal Covariate Shift," ArXiv, vol. abs/1502.03167, 2015.

[9] H. Ismail Fawaz, G. Forestier, J. Weber, L. Idoumghar, and P.-A. Muller, "Deep learning for time series classification: a review," Data Mining and Knowledge Discovery, vol. 33, no. 4, pp. 917-963, 2019/07/01 2019, doi: 10.1007/s10618-019-00619-1.